\pdfoutput=1
\documentclass{IEEEoj}
\usepackage{cite}
\usepackage{amsmath,amssymb,amsfonts}
\usepackage{algorithmic}
\usepackage{graphicx,color}
\usepackage[caption=false]{subfig}
\usepackage{textcomp}
\usepackage{array}
\usepackage{hyperref}

\def\BibTeX{{\rm B\kern-.05em{\sc i\kern-.025em b}\kern-.08em
    T\kern-.1667em\lower.7ex\hbox{E}\kern-.125emX}}
\def\OJlogo{\vspace{-14pt}\includegraphics[height=28pt]{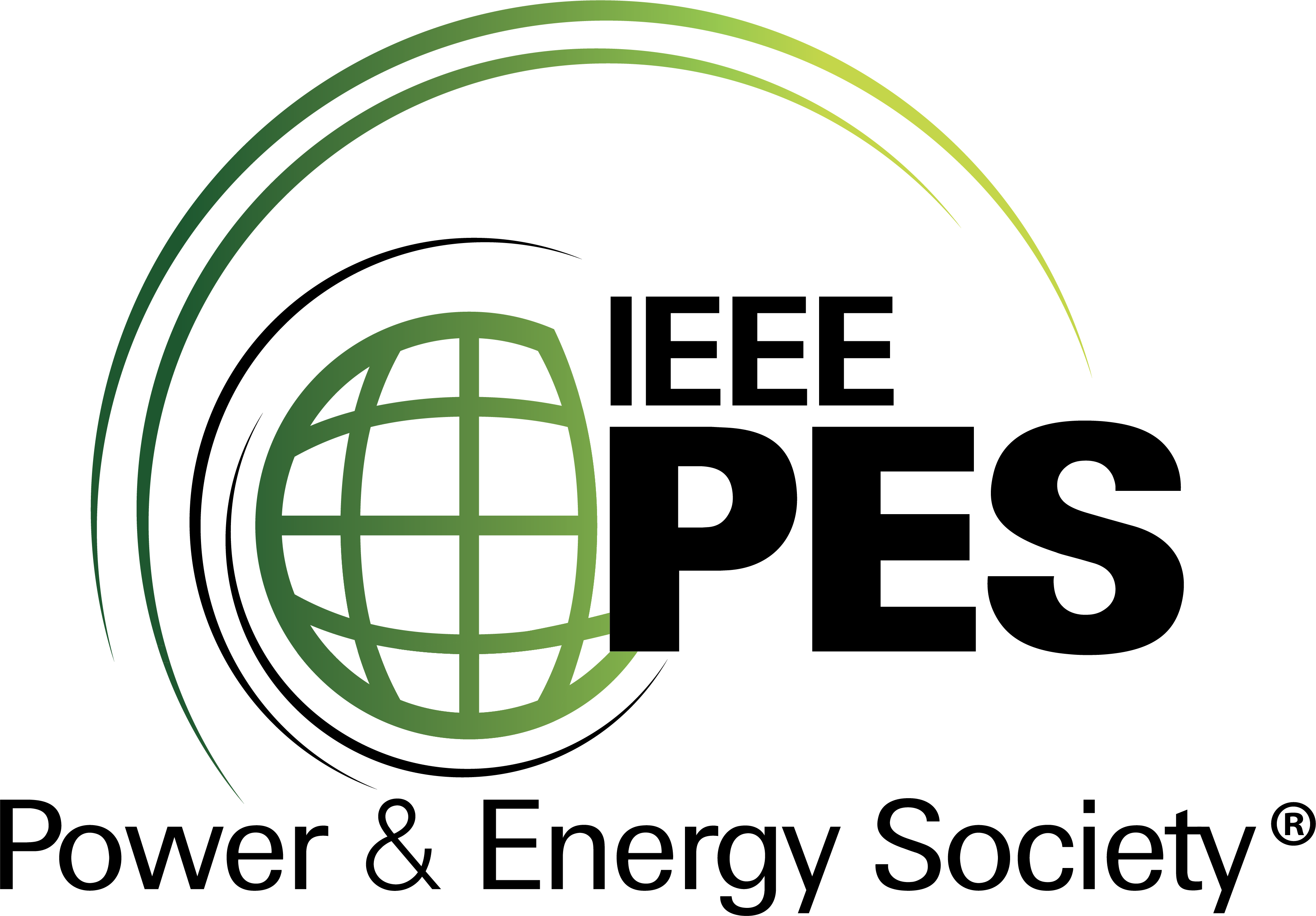}}
\begin{document}


\title{EPIGRIDS: Documentation and Validation of Seven Open Synthetic Transmission Grid Cases}

\author{JONATHAN M. SNODGRASS\affilmark{1}, SENIOR MEMBER, IEEE,
SCOTT GREENE\affilmark{2},
THOMAS J. OVERBYE\affilmark{1}, LIFE FELLOW, IEEE,
CHRISTOPHER L. DeMARCO\affilmark{2}, LIFE MEMBER, IEEE, AND
BERNARD C. LESIEUTRE\affilmark{2}, FELLOW, IEEE}
\affil{Department of Electrical and Computer Engineering, Texas A\&M University, College Station, TX 77843 USA}
\affil{Department of Electrical and Computer Engineering, University of Wisconsin-Madison, Madison, WI 53706 USA}
\corresp{CORRESPONDING AUTHOR: Jonathan M. Snodgrass (e-mail: Snodgrass@tamu.edu).}
\authornote{This work was supported in part by the U.S. Department of Energy, Advanced Research Projects Agency-Energy (ARPA-E), under the GRID DATA program, award DE-AR0000717, through the EPIGRIDS (Electric Power Infrastructure and Grid Representations in Interoperable Data Sets) project.}
\markboth{EPIGRIDS: Seven Open Synthetic Transmission Grid Cases}{Snodgrass \textit{et al.}}

\begin{abstract}
This paper documents and checks EPIGRIDS, seven open synthetic transmission
grid cases from four independently built networks, made with an
optimization-based construction method instead of the
graph-statistics method behind most large public synthetic grids. The cases range from 250 to 78{,}484 buses, with geolocated substations, voltages up to
765~kV, and line parameters taken from conductor data. Each ships under an open
license in three common power flow formats that agree at the solved operating
point, and the six larger cases have a companion hourly, bus-level load series
for one year. The Texas, Florida, and
Michigan, Ohio, and Indiana cases are sub-networks of the Eastern case. Every
case solves an alternating-current power flow from its stored operating point.
With parallel circuits merged, the Eastern and Western cases are 9 and 16\,\%
denser than the real interconnections, while graph-statistics cases on the
same footprints lie within 6\,\% and meet more published realism criteria. A full single-outage contingency
analysis completes on every case, and in an optimal power flow benchmark every
case except the New England base case and one of its 12 operating points
solves in polar form.
\end{abstract}


\maketitle

\section{Introduction}\label{sec:intro}

\IEEEPARstart{R}{esearchers} who develop power flow, contingency, and optimal
power flow (OPF) methods, and the artificial intelligence (AI) models now
trained to approximate them, need realistic transmission networks that can be
shared. Detailed models of the U.S. grid are critical energy infrastructure
information (CEII) and cannot be published (Section~\ref{sec:bg}), so the
community relies on synthetic grids. What is lacking is
variety in the methods used to build them. Most large public synthetic
transmission cases come from a single family of methods
\cite{Birchfield_2017_GridStructuralCharacteristics}, and the datasets used to
train and benchmark machine learning for OPF are built by perturbing a handful
of base cases, mostly IEEE test systems and other PGLib-OPF networks
\cite{Lovett_2024_OPFData, Varbella_2024_PowerGraph,
Klamkin_2025_PGLearn, Puech_2025_GridfmDatakit}. Graph neural networks, OPF
surrogates, and grid foundation models have shown promise for learning fast
approximations of power system computations
\cite{Hamann_2024_FoundationModelsGrid}, and whether such a model works on grids
built by a different method is an open question.

There are three broad strategies for producing ``realistic but not real'' grids
without exposing CEII: building fictitious networks from scratch, obfuscating a
real grid, and reconstructing an approximate model of the real grid from open
geographic data (Section~\ref{sec:bg}). Each can leave its own structural and
electrical signature in the grids it produces. Since one method dominates the
built-from-scratch cases most used for training and benchmarking, a second,
independently derived family of built-from-scratch grids adds directly to the
variety of construction methods.

This paper documents such a family, EPIGRIDS (Electric Power Infrastructure and
Grid Representations in Interoperable Data Sets). Its cases were built with
the ``KCL-only'' construction code, an
optimization-based method that selects transmission corridors with a
network-flow linear program instead of growing a network to match graph
statistics \cite{Snodgrass_2021_Dissertation, Snodgrass_2026_GridForge}. The
seven cases are released openly, alongside the Birchfield and Overbye cases, on
the Texas A\&M Electric Grid Test Case Repository
\cite{Kunkolienkar_2024_TestCaseRepo}. They span more than two orders of magnitude
in size, from a 250-bus New England case to a 78{,}484-bus model of the full
U.S. Eastern Interconnection footprint, and cover single-state, multi-state,
and interconnection-scale footprints (Table~\ref{tab:cases}).

The cases themselves are not new. They have been public since 2022
\cite{WISPOPOP_2022_SyntheticModels}, five of them have been in PGLib-OPF since
v23.07 \cite{PGLibOPF_2023_v2307}, and the
construction method is presented in \cite{Snodgrass_2021_Dissertation,
Snodgrass_2026_GridForge}. What has been missing is a citable description of the cases as released:
\cite{Snodgrass_2026_GridForge} presents the method and evaluates it on its
regional networks, while this paper documents the seven released AC cases
themselves. This paper provides it in five parts: 1)~a documented
release of the seven CEII-free cases in three formats, under the CC~BY~4.0
license with a DOI; 2)~a build history of each case, reconstructed from the
surviving files, which documents that the Texas, Florida, and
Michigan, Ohio, and Indiana (MIOHIN) cases are sub-networks of the Eastern case;
3)~validation of the cases as released, including AC power flow, a structural
comparison with the real Eastern and Western Interconnections and with
same-footprint graph-statistics cases, and a full AC \mbox{$N\!-\!1$}
contingency analysis of every case; 4)~an OPF benchmark of all seven cases
across seven PowerModels.jl formulations, positioned against PGLib-OPF v23.07;
and 5)~a companion 8{,}760-hour, bus-level load series mapped to six of the
cases.

Section~\ref{sec:bg} reviews the construction strategies and places EPIGRIDS
among them. Section~\ref{sec:cases} describes the seven released cases,
Section~\ref{sec:construction} how they were built, Section~\ref{sec:validation}
their validation, and Section~\ref{sec:studies} the benchmark studies, limitations, and
intended use; Section~\ref{sec:conc} concludes.

\section{Background and Related Work}\label{sec:bg}

Realistic large-scale grid models are not public because of security policy.
After September~11, 2001, the Federal Energy Regulatory Commission (FERC)
designated detailed models of the U.S. grid as CEII and withdrew them
from public release, along with the transmission data utilities file through
Form~715 \cite{FERC_2018_Form715}.
The community's response,
accelerated by the ARPA-E GRID DATA program, has been to build ``synthetic''
grids: fictitious systems that are statistically and electrically realistic,
contain no CEII, and can be shared freely \cite{ARPAE_2018_GridData,
Aravena_2025_OpenPowerSystemDatasets}. Table~\ref{tab:compare} compares
representative families.

\subsection{Synthetic Grids Built From Scratch}\label{subsec:scratch}
The dominant approach, the method of Birchfield, Gegner, Xu, Shetye, and
Overbye, builds a network from public geographic and statistical data. It places substations
and load realistically \cite{Gegner_2016_GeographicallyRealistic}, grows a
transmission network tuned to validation criteria from real systems
\cite{Birchfield_2017_GridStructuralCharacteristics,
Birchfield_2017_MetricBasedValidation}, and refines the result until it
supports a converged, reactively planned AC power flow across many voltage
levels \cite{Birchfield_2018_BuildingManyVoltageLevels,
Birchfield_2018_PowerFlowConvergence}. Later work added time-varying load \cite{Li_2021_LoadTimeSeries}. The resulting ACTIVSg family
is a widely used body of open synthetic grids
\cite{Kunkolienkar_2024_TestCaseRepo}, and the U.S. test system of
\cite{Xu_2020_USTestSystem} pairs a Texas A\&M synthetic network with load and
renewable time series. Other generators grow random or spatially embedded
topologies that match the statistics of real networks
\cite{Wang_2010_RandomTopologies, Elyas_2017_CorrelatedBusTypes,
Soltan_2015_SpatiallyEmbedded}.

A second, less widely known line of work treats construction as an
optimization problem: it selects network structure and parameters subject to
power flow physics \cite{Thiam_DeMarco_2014_TransmissionExpansion}. EPIGRIDS belongs to this line;
Section~\ref{sec:validation}-\ref{subsec:realism} measures how its
interconnection cases differ in structure from graph-statistics cases on the
same footprints.

\subsection{Datasets Derived From Real Grids and Open-Data Reconstructions}\label{subsec:obfuscated}
The alternative strategy starts from a real network and removes its
confidential character. The pan-European PEGASE cases reach 13{,}659 buses, and
snapshots of the French grid from 2013 reach 6{,}515 buses
\cite{Josz_2016_ACPowerFlowData}. Later work obfuscates the line and
transformer parameters of a network with differential privacy while keeping it
AC feasible \cite{Fioretto_2020_DPObfuscation}, and the Sustainable Data Evolution
Technology (SDET) of Pacific Northwest National Laboratory (PNNL) learns
statistical features from proprietary industry datasets \cite{Huang_2018_SDET}.
A third strategy reconstructs an approximate model of the real grid from open
geographic data such as OpenStreetMap \cite{Britto_2026_OpenDataGridModels}, and open models such as PyPSA-Eur \cite{Horsch_2018_PyPSAEur} pair such
networks with load and renewable time series.
These datasets inherit the realism of their source networks, but each reflects
the quirks of one source system and region, so a built-from-scratch family
complements them.

\subsection{Repositories, Benchmarks, and Validation}\label{subsec:repos}
Repositories such as the Texas A\&M Electric Grid Test Case Repository
\cite{Kunkolienkar_2024_TestCaseRepo} collect these cases, and a curated subset
forms the Power Grid Library (PGLib-OPF), used to benchmark AC OPF algorithms
\cite{PGLibOPF_2023_v2307, Babaeinejadsarookolaee_2019_PGLib}. The ARPA-E
Grid Optimization Competition released large security-constrained OPF
datasets \cite{Aravena_2022_GOCompetition}. Most
machine-learning datasets and dataset toolkits for OPF, such as OPFData
\cite{Lovett_2024_OPFData}, PowerGraph \cite{Varbella_2024_PowerGraph}, PGLearn
\cite{Klamkin_2025_PGLearn}, and gridfm-datakit \cite{Puech_2025_GridfmDatakit},
build their samples by perturbing this handful of base networks. The number of independent base grids, and of the
methods used to build them, therefore matters for downstream learning.

\begin{table*}[!t]
\centering
\caption{Representative families of open synthetic transmission-grid datasets.
EPIGRIDS is a second built-from-scratch family, complementing the widely used
graph-statistics cases.}
\label{tab:compare}
\begin{tabular}{@{}llllccl@{}}
\hline
\textbf{Dataset (method)} & \textbf{Strategy} & \textbf{Source grid} &
\textbf{Scale (buses)} & \textbf{Geolocated} & \textbf{Hourly load} &
\textbf{License}\\
\hline
ACTIVSg (Birchfield-Overbye) & from scratch & none (fictitious) & $\sim$200--82{,}000 & yes & yes & CC BY 4.0\\
\textbf{EPIGRIDS (KCL-only)} & from scratch & none (fictitious) & 250--78{,}484 & yes & yes & CC BY 4.0\\
PEGASE & obfuscation & Europe (data altered) & up to 13{,}659 & no & no & CC BY 4.0\\
RTE snapshots & obfuscation & France (2013) & 1{,}888--6{,}515 & no & no & CC BY 4.0\\
\hline
\end{tabular}
\end{table*}

\section{The EPIGRIDS Cases}\label{sec:cases}
This section describes the seven released cases as a user receives them. They
were developed under the ARPA-E GRID DATA program \cite{ARPAE_2018_GridData},
published in the
\emph{Synthetic Electric Network Models} repository
\cite{WISPOPOP_2022_SyntheticModels}, and listed on the Texas A\&M Electric
Grid Test Case Repository \cite{Kunkolienkar_2024_TestCaseRepo}. Three versions of one construction code, called ``KCL-only'' for its
network-flow model, which enforces Kirchhoff's
current law (KCL) but not the voltage law \cite{Snodgrass_2021_Dissertation,
Snodgrass_2026_GridForge}, produced them (Section~\ref{sec:construction}), and
Table~\ref{tab:cases} summarizes each case.

\subsection{Coverage and Scale}\label{subsec:coverage}
The release holds two interconnection-scale cases, three regional cases cut
from one of them, and two smaller cases. The Eastern case covers 37 states and
the District of Columbia in the Eastern Interconnection footprint at their full
geographic extent, so all of Texas lies inside it, and the Western case covers 11 states. The Texas case is therefore a
geographic cut of the Eastern case, synchronous with it, and not a model of the
ERCOT Interconnection. The MIOHIN
(Michigan, Ohio, and Indiana; release folder ``Midwest''), Texas, and Florida
cases are sub-networks of the Eastern case (Section~\ref{sec:construction}-\ref{subsec:extract}).
Wisconsin\_1664 covers one state and New England~250 covers six, although four
of its buses lie in New York.

Each region was built with the two or three high voltages (230~kV and above)
that its utilities use most, plus 138 and 69~kV subtransmission
\cite[Ch.~3]{Snodgrass_2021_Dissertation}. The highest voltage is therefore
regional, as it is in practice: 345~kV in New England, Wisconsin, and Texas,
500~kV in Florida and Western, and 765~kV in MIOHIN and Eastern.

\begin{table*}[!t]
\centering
\caption{The seven released EPIGRIDS cases.}
\label{tab:cases}
\footnotesize
\setlength{\tabcolsep}{4pt}
\begin{tabular}{@{}lccccccc@{}}
\hline
& \textbf{Eastern} & \textbf{Western} & \textbf{MIOHIN} & \textbf{Texas} &
\textbf{Florida} & \textbf{Wisconsin} & \textbf{New England}\\
\hline
Footprint               & 37 states, DC & 11 states & MI, OH, IN & TX & FL & WI & 6 states\\
Buses                   & 78{,}484 & 20{,}758 & 10{,}192 & 7{,}336 & 5{,}658 & 1{,}664 & 250\\
Lines                   & 88{,}429 & 24{,}858 & 12{,}053 & 8{,}058 & 6{,}431 & 1{,}902 & 248\\
Transformers            & 37{,}717 & 8{,}510 & 4{,}990 & 3{,}463 & 2{,}647 & 560 & 91\\
Units (in service)      & 6{,}873 (6{,}773) & 2{,}250 (2{,}174) & 722 (714) & 686 (684) & 474 (474) & 79 (79) & 42 (42)\\
Installed capacity (GW) & 863.5 & 210.4\textsuperscript{a} & 92.4 & 111.8 & 66.7 & 18.5 & 29.8\\
Load, base case (GW)    & 515.0 & 120.9 & 76.5 & 66.3 & 42.4 & 10.0 & 11.3\\
Voltage levels (kV)     &
\begin{tabular}[t]{@{}c@{}}765, 500, 345,\\ 230, 138, 69\end{tabular} &
\begin{tabular}[t]{@{}c@{}}500, 345, 230,\\ 138, 69\end{tabular} &
\begin{tabular}[t]{@{}c@{}}765, 500\textsuperscript{b}, 345,\\ 138, 69\end{tabular} &
\begin{tabular}[t]{@{}c@{}}345, 230,\\ 138, 69\end{tabular} &
\begin{tabular}[t]{@{}c@{}}500, 230,\\ 138, 69\end{tabular} &
\begin{tabular}[t]{@{}c@{}}345, 230,\\ 138, 69\end{tabular} &
\begin{tabular}[t]{@{}c@{}}345, 230, 138,\\ 69, 14\textsuperscript{b}\end{tabular}\\
Line length (km)\textsuperscript{c} & 723{,}585 & 224{,}558 & 84{,}777 & 71{,}455 & 37{,}855 & 18{,}701 & 7{,}636\\
Operating points        & 1 & 1 & 1 & 1 & 1 & 1 & 13\\
\hline
\end{tabular}
\par\vspace{3pt}
\parbox{\textwidth}{\footnotesize \textsuperscript{a}From the .RAW and .pwb;
the .m gives 207.1~GW (Section~\ref{sec:cases}-\ref{subsec:formats}).
\textsuperscript{b}One isolated 500~kV bus in MIOHIN; nine 14~kV generator buses in New
England. \textsuperscript{c}Sum of the great-circle distances between the end
buses of each line.}
\end{table*}

\subsection{Model Contents}\label{subsec:contents}
Every case is a geolocated, multi-voltage transmission model. In the five
largest cases and in Wisconsin, synthetic substations sit at the centroids of
U.S. census tracts, so bus density follows population and load, and each
substation holds one bus per voltage level joined by transformers
\cite{Snodgrass_2021_Dissertation}. Lines and transformers carry resistance,
reactance, charging susceptance, a base-case rating (rating~A), and a
post-contingency rating (rating~C). Branch loading in this paper is
PowerWorld's percent loading computed from current (its Amp/MVA basis).
Generating units carry fuel and unit-type labels (in all cases but New
England) and polynomial costs of at most second order, derived from typical
heat rates and fuel costs in a report included with the release
\cite{Babaeinejadsarookolaee_2018_CostCurves}. As released, 286 Eastern units,
89 of them nuclear, have zero cost. The cases also hold switched shunts (from
none in Florida to 452 in Western), areas by state in Eastern, Western, and
MIOHIN, and, in Eastern and Texas, a 550~MW two-terminal DC line with both ends
in Texas.

\begin{figure*}[!t]
\centering
\subfloat[New England, 250 buses]{%
  \includegraphics[height=1.7in]{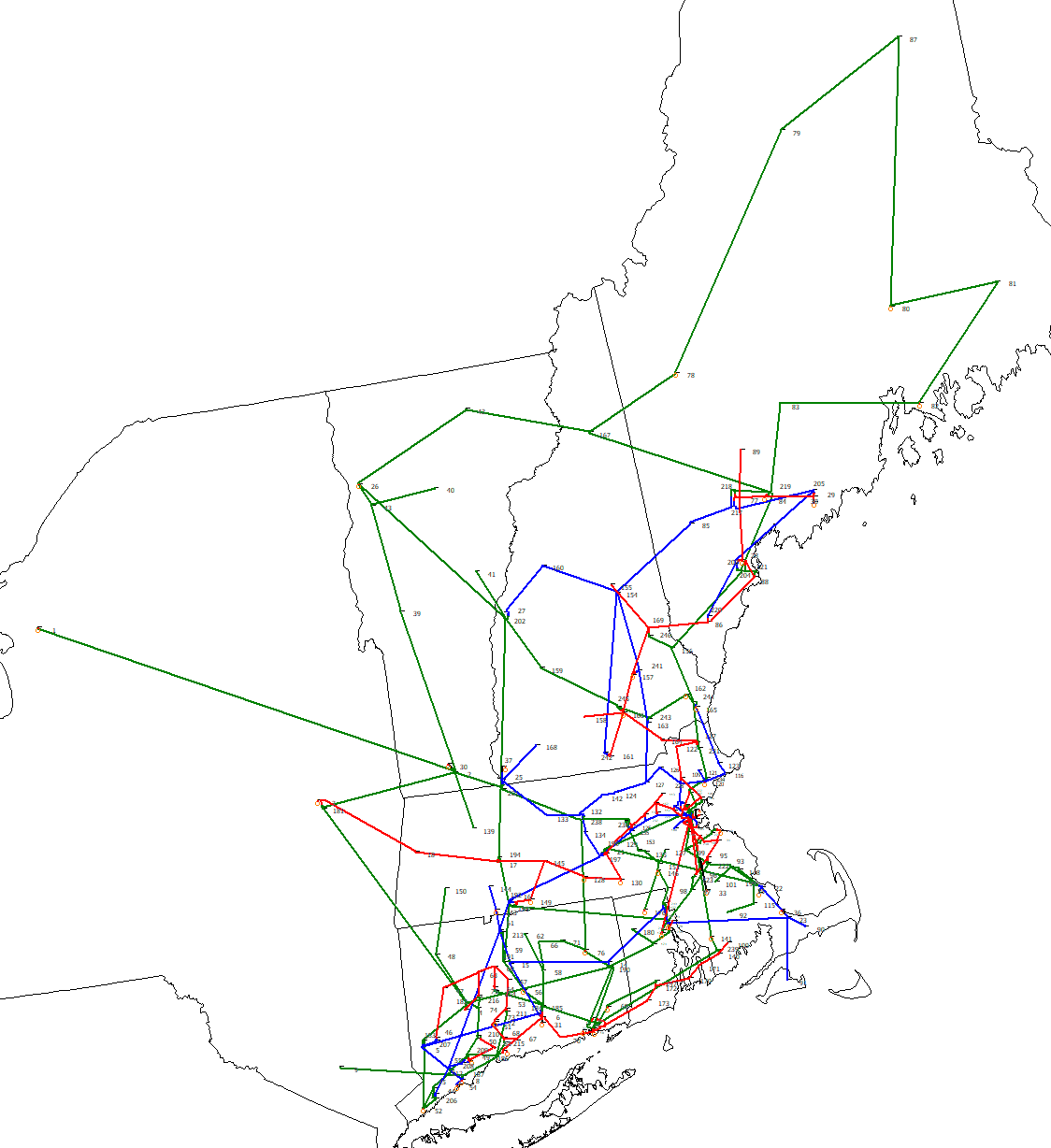}%
  \label{fig:onelines-ne}}\hfil
\subfloat[Wisconsin, 1{,}664 buses]{%
  \includegraphics[height=1.7in]{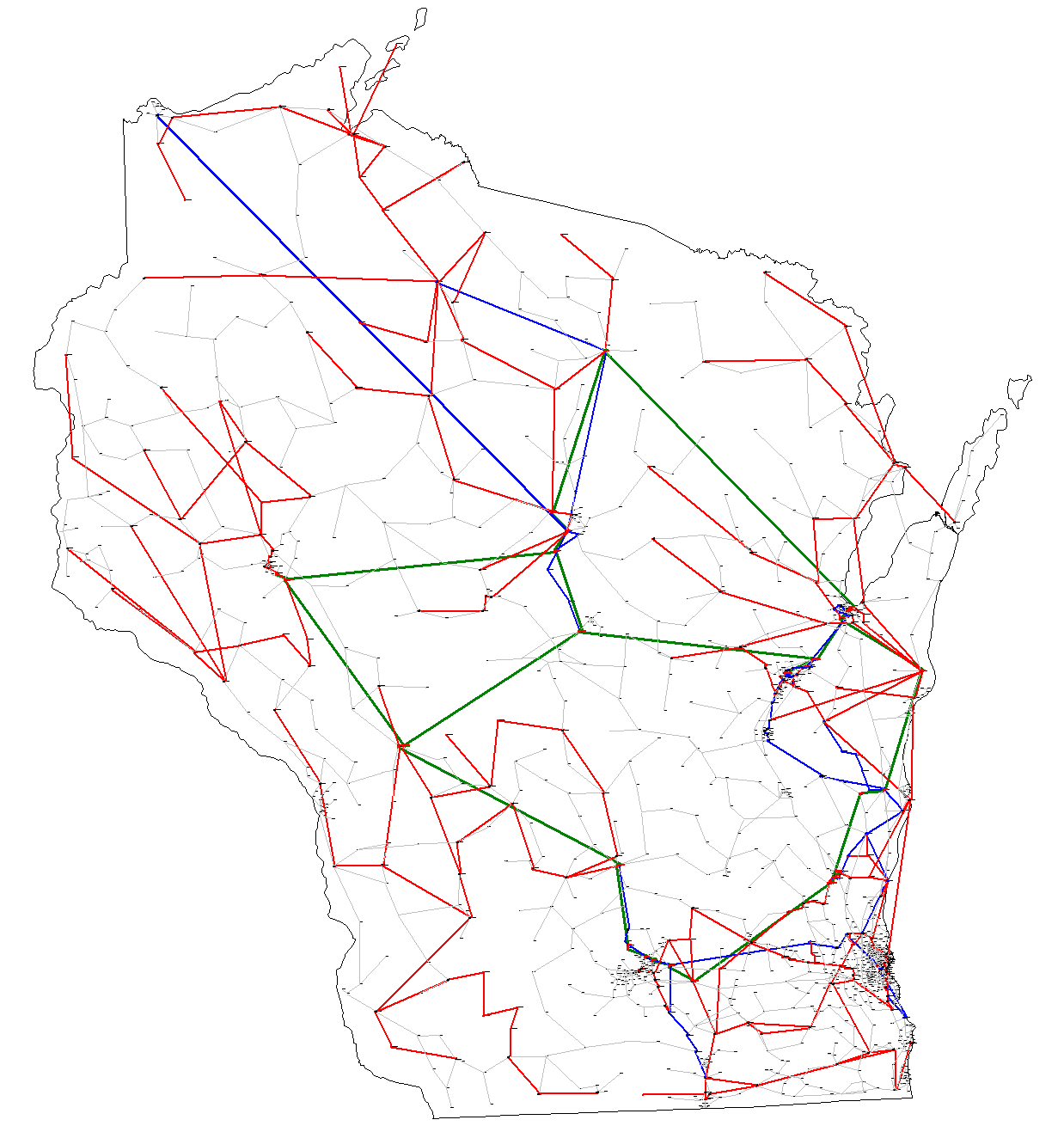}%
  \label{fig:onelines-wi}}\hfil
\subfloat[Western, 20{,}758 buses]{%
  \includegraphics[height=1.7in]{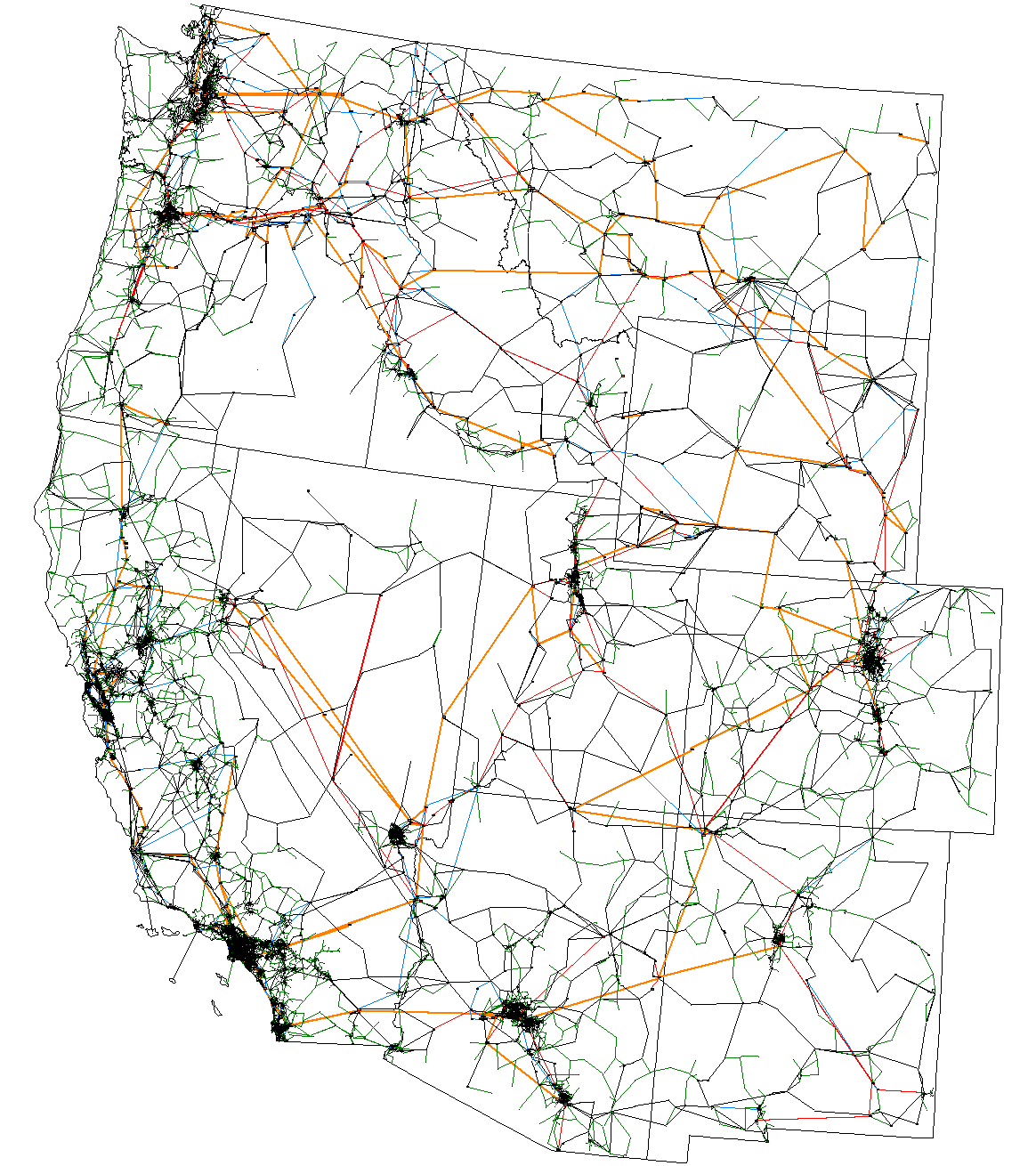}%
  \label{fig:onelines-west}}\hfil
\subfloat[Eastern, 78{,}484 buses]{%
  \includegraphics[height=1.7in]{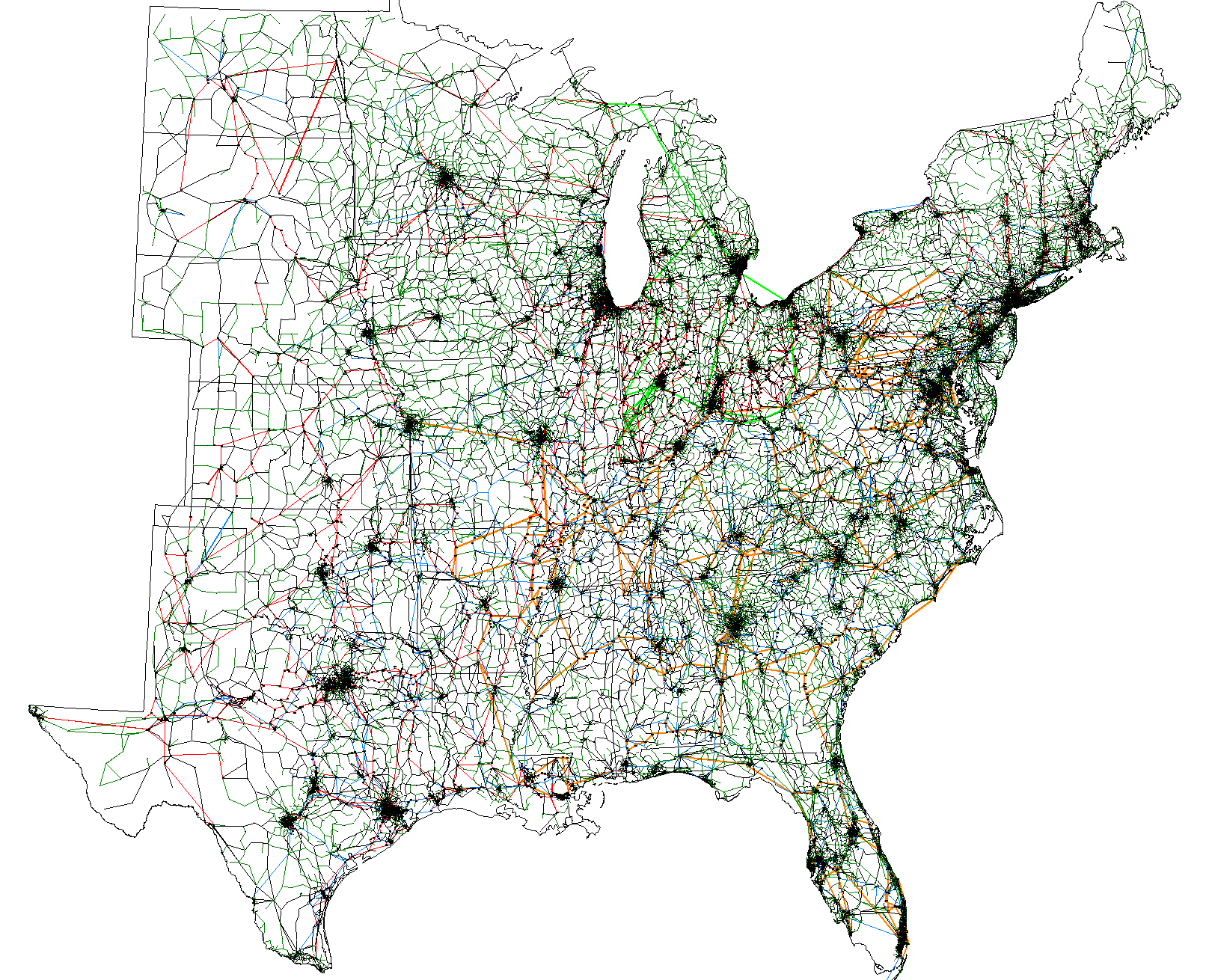}%
  \label{fig:onelines-east}}
\caption{One-line diagrams of four released cases, covering the three build families
(Section~\ref{sec:construction}-\ref{subsec:families}): (a) New England, family~A; (b) Wisconsin,
family~B; (c) Western and (d) Eastern, family~C. Texas, Florida, and MIOHIN are
sub-networks of (d). Line colors give the voltage: in (a) and (b), grey 69~kV,
red 138~kV, blue 230~kV, and dark green 345~kV; in (c) and (d), black 69~kV,
dark green 138~kV, light blue 230~kV, red 345~kV, orange 500~kV, and bright
green 765~kV. Thin black outlines are state borders. One-lines of the same
Wisconsin, Western, and Eastern cases appear in \cite{Snodgrass_2026_GridForge}.}
\label{fig:onelines}
\end{figure*}

In the five largest cases, line parameters trace to conductors that could be
built rather than to sampled impedance distributions. For each synthetic
region, a lookup table spans the conductor sizes and bundles that utilities
report in FERC Form~1 at each voltage. Each entry pairs an MVA rating from
ampacity data \cite{Southwire_2012_ACSR} with resistance and reactance terms
from \cite{LaForest_1982_TransmissionLineReference} and a phase spacing typical
of the region and voltage \cite[Ch.~3]{Snodgrass_2021_Dissertation},
\cite{Snodgrass_2026_TransmissionLineDesign}. Fig.~\ref{fig:onelines} shows four cases that cover the three build
families and contain all seven released networks.

\subsection{Formats and Their Agreement}\label{subsec:formats}
Each case ships in three formats, all written by PowerWorld Simulator. They
are a Simulator case (.pwb) with its one-line diagram (.pwd)
\cite{PowerWorld_2026_Simulator}, a MATPOWER case (.m, format version~2)
\cite{Zimmerman_2011_MATPOWER}, and a PSS/E raw data file (.RAW; revision~36
for the six larger cases, revision~33 for New England) \cite{Siemens_PSSE_RAW}. Compared element by element, the three formats hold the same network and
operating state, in the 2023 release files and in the 2026 re-export of the six
larger cases (Simulator version~24, PSS/E revision~36). Counts and total load agree exactly, unit outputs agree within 0.005~MW, and bus voltages agree
within 0.00001~pu at every connected bus.

Apart from bus-type codes and the voltages of isolated buses (six in Eastern,
three in MIOHIN), every difference comes from how Simulator writes a MATPOWER
file. Units off automatic generation control are written with
$P_{\min}=P_{\max}=P_g$, so the Western .m understates installed capacity by
3.3~GW. Units off voltage control are written with $Q_{\min}=Q_{\max}=Q_g$, and
load Mvar, modeled as constant impedance in the .pwb and .RAW, is written as
constant power at the solved voltage. A MATPOWER user therefore solves a
slightly different model: an AC power flow of the Texas .m without reactive
limits moves bus voltages by up to 0.06~pu and overloads 21 branches; the .RAW
and .pwb hold the true unit limits.

\subsection{Operating Points and the Companion Load Series}\label{subsec:load}
Each of the six larger cases ships one solved operating point, and its loads
record how the case was built. The Eastern loads equal hour 5366 (August~12,
2015, hour ending~14) of the synthetic load series
\cite{Snodgrass_2026_SyntheticLoad}, assembled state by state from different
versions of the series. The Western loads equal hour 5033 (July~29, 2015,
hour ending~17). Texas, Florida, and MIOHIN carry the Eastern loads on the same
buses, and the Wisconsin loads are 0.70 times those of a 2018 Wisconsin case
with 14.3~GW of load. New England ships a base case and 12 operating points in Easy, Medium, and
Hard folders (13 .m and .RAW files; 20 .pwb files, with nine operating points under each
voltage-limit set), under
``Generous'' and ``Tight'' voltage-limit sets (load buses 0.92 to 1.06 and
0.95 to 1.04~pu). The number in each file name is an hour of a 168-hour seasonal week of
2010 load, with total loads of 9{,}390 to 24{,}044~MW
across the files; the base case (11{,}324~MW) is
hour~168 of the fall week.

The companion load dataset \cite{Snodgrass_2026_LoadSmithZenodo} is the 2026
output of LoadSmith, the Python version
of the synthetic load code, with peak scaling set to ``none'', so that its hourly loads are not
rescaled to a regional peak target. It gives
8{,}760 hours of real and reactive power for 72{,}066 load buses in 13 regions
and maps by census-tract name to 95.9\,\% (Texas) to 100\,\% of the load MW of
the six larger cases. Its method is that of \cite{Snodgrass_2026_SyntheticLoad}
with four differences in the code: higher power-factor ranges, no smoothing, an
NREL load basis in 29 of 49 state fits, and
California load from PG\&E alone.

\subsection{License, Access, and Release Fixes}\label{subsec:access}
The cases are released under the Creative Commons Attribution 4.0
International license (CC~BY~4.0), copyright 2022 J. Snodgrass, S. Greene,
B. Lesieutre, and C. DeMarco. They were built from public data and from aggregate statistics: transformer
impedances follow the statistics of the 2010 Eastern Interconnection MMWG
planning case, and no facility-level data entered the build. The real-network values reported in
Section~\ref{sec:validation} are aggregate statistics (node-degree fit areas and
voltage-class shares) that identify no facility. The version described here is archived on Zenodo from a
tagged GitHub release \cite{Snodgrass_2026_EPIGRIDSZenodo}. With this paper,
the release documentation was corrected (voltage lists, MATPOWER export
conventions, isolated buses, New England file labels) and extended with bus
counts by voltage, capacity by fuel, and the New England operating points.

\section{Construction and Build History}\label{sec:construction}
This section documents how each released case was built, so that a user
knows which cases share a network and which steps rest on surviving files. We
compared the dated intermediate networks of 2017 to 2020 with the released
files, step by step; the method itself is described in
\cite{Snodgrass_2021_Dissertation, Snodgrass_2026_GridForge}. Each step carries a confidence: ``verified'' when an intermediate file matches the
release, ``inferred'' when a rule reproduces the released values, and ``values
only'' when we know the change but found no rule.

\subsection{Three Build Families}\label{subsec:families}
All seven cases come from one line of KCL-only construction code (its
network-flow model enforces Kirchhoff's current law but not the voltage law), developed
from 2017 to 2019 \cite[Ch.~7]{Snodgrass_2021_Dissertation}, and fall into three families by
code version. Family~A is New England~250 (2017, earliest version). Family~B is
Wisconsin\_1664 (2017 and 2018, 1{,}393 substations, earlier voltage and
transformer steps). Family~C is the national code of 2018 and 2019, which built
the Western and Eastern cases in 2019. Texas, Florida, and MIOHIN were cut from
the Eastern case in January 2020. The New
England and Wisconsin networks, the joining of both interconnection cases, and
the regional cuts are verified against surviving files; the Western impedance
change is inferred; and the PowerWorld edits of November 2019 to January 2020
are known by their values only (Section~\ref{sec:construction}-\ref{subsec:finishing}).

\subsection{The KCL-Only Construction Method in Brief}\label{subsec:kclonly}
The KCL-only method builds a network in two stages
\cite[Chs.~4 and~7]{Snodgrass_2021_Dissertation}, \cite{Snodgrass_2026_GridForge}.
The first stage chooses the corridors. Its inner loop solves a network-flow
linear program (LP) that minimizes the weighted $L_1$ norm of the branch flows
subject to KCL and the limits of the existing lines:
\begin{equation}\label{eq:kcl-lp}
\begin{aligned}
\min_{\mathbf{p}}\quad & \textstyle\sum_{j} c_j \lvert p_j\rvert\\
\text{s.t.}\quad & \mathbf{A}\mathbf{p} = \mathbf{P},
  \qquad \textstyle\sum_{i} P_i = 0,\\
& \lvert p_j\rvert \le \bar{p}_j \quad \forall\, j \in \mathcal{E},
\end{aligned}
\end{equation}
where $\mathbf{p}$ is the branch-flow vector, $\mathbf{A}$ the bus-branch incidence matrix of the existing lines $\mathcal{E}$ and the candidate
corridors $\mathcal{C}$, $\mathbf{P}$
the net injections, $c_j$ a weight proportional to the length of branch $j$,
and $\bar{p}_j$ the MVA rating of existing line $j \in \mathcal{E}$, used
as a limit on its real power flow. The LP follows the transportation model of
Garver \cite{Garver_1970_TransmissionLP}; the weights are those of
\cite[Ch.~4]{Snodgrass_2021_Dissertation}: 1 per unit length for existing lines, 5 for
new corridors, and 3 for candidates parallel to an existing line. Because an
$L_1$ objective gives a sparse solution, the flow on most candidates is zero to
numerical precision. A middle loop adds the highest-flow candidates as lines, rated from
their flow, until the LP adds none. An outer loop repeats this for an economic dispatch, a dispatch that runs the
units left below 85\,\% of capacity at full output, and the loss of each
generating substation.

The second stage assigns parameters. Each line takes the conductor from the
regional table (Section~\ref{sec:cases}-\ref{subsec:contents}) that matches its
rating, a sequential DC power flow resizes overloaded and underused lines, and each
substation gets one bus per voltage level. Transformers take their ratings
from the connected capacity and their impedances from statistics of the 2010
Eastern Interconnection MMWG planning case. Interconnection-scale cases are
built from 13 regional networks, so that every LP stays tractable.

\subsection{Finishing the Interconnection Cases}\label{subsec:finishing}
Neither interconnection case came from one LP; both were joined from their
regional networks with the reconnection step of
Section~\ref{sec:construction}-\ref{subsec:kclonly}. The Western case joined
three networks (the Pacific Northwest, California, and the Desert Southwest) in
September 2019, and the Eastern case joined ten, holding 58{,}574 substations,
in October 2019. A second round of the LP then reconnected each Eastern voltage layer in just
under 22 hours. Of the 11{,}382 Eastern corridors that no regional network
contains, this round created 11{,}341 (99.6\,\%) and hand edits added the other
41. Re-sizing and transformers followed.

PowerWorld finishing ran from November 2019 to January 2020 in five steps.
First, loads came from the synthetic load mapping
(Section~\ref{sec:cases}-\ref{subsec:load}). In the Eastern case, 1{,}708 loads
(71{,}883~MW) moved from high-voltage buses to the 138 or 69~kV bus of their
substation. Second, Western line resistance and reactance became 1/1.4 of their
2019 values on 96 to 97\,\% of lines, which removes the route factor the 2019 code applied to line length
(inferred). The released Eastern and Western lines have the same reactance per
mile within 2\,\% at 138 to 500~kV, so neither keeps it. Third, hand edits gave the Eastern case
13 line-splitting buses and 41 corridors, and the Western case 661 parallel
circuits, 34 corridors, and four 500~kV midpoint buses. Fourth, control data
went in: switched shunts, voltage setpoints, a slack bus, areas by state, and the 550~MW DC
line in Eastern. Fifth, one of the authors (S. Greene) and undergraduate assistants brought
each case to a solved AC operating point and made the edits above; no intermediate files of
this step survive, so it is known by its result only.

\subsection{Regional Cases as Induced Sub-Networks}\label{subsec:extract}
Texas, Florida, and MIOHIN are induced sub-networks of the Eastern case: each
holds its buses, its branches (with the same resistance, reactance, charging,
and rating~A), and its units exactly as in the Eastern case. The extraction
dropped the ties to the rest of the Eastern case (69 for Texas, 41 for Florida,
and 75 for MIOHIN), gave Florida and MIOHIN new slack buses, and redispatched generation by
participation factor;
we verified each rule against the Eastern case file. A user should therefore treat the three as views of one build and
one load snapshot, not as independent networks.

\section{Validation and Quality Assurance}\label{sec:validation}
This section checks the cases as released. The checks ran in October 2026 with the tools listed in
Section~\ref{sec:validation}-\ref{subsec:repro}, on the case files archived as
release v1.0.0 \cite{Snodgrass_2026_EPIGRIDSZenodo}; the .m files there are
identical to the 2023 release files.

\subsection{Internal Consistency and Solvability}\label{subsec:solve}
Section~\ref{sec:cases}-\ref{subsec:formats} showed that the three formats of
each case hold the same network and operating state. Each of the 26 .pwb files
(the six larger cases and the 20 New England cases) solves an AC power flow in
PowerWorld from its stored operating point. An independent Newton power
flow in PYPOWER \cite{Lincoln_PYPOWER} on the .m files, with reactive limits
enforced, agrees with PowerWorld within 0.0075~pu at every bus of every case
except MIOHIN, where the largest difference is 0.035~pu
(Table~\ref{tab:validation}). In one 120-bus pocket, the two tools leave
different units at their reactive limits, and imposing PowerWorld's limit
statuses on 36 units there brings PYPOWER within 0.0001~pu. Solved voltages stay between 0.904 and
1.090~pu. Three cases carry branches above 100\,\% of rating~A in the base
case: one in Eastern, 10 in MIOHIN (up to 140.7\,\%), and two in Wisconsin (up
to 126.2\,\%). The Eastern branch, 20422-21190 \#1, loads at 100.0\,\% on
PowerWorld's Amp/MVA basis but at 97.9\,\% on an MVA basis, so the basis
matters at the margin. Seven Eastern and 9 MIOHIN 69~kV lines that the
release files leave unmonitored exceed even their rating~C in the base case.
Many units sit at a reactive limit in the solved cases: in Eastern, 3{,}617 are at
their maximum and 257 at their minimum. By a
common test on the .m files (reactive output within 0.1~Mvar of the maximum),
53.7\,\% of in-service Eastern units sit at their maximum, against 36.5\,\% in
the synthetic ACTIVSg70k case of the same footprint.

\begin{table*}[!t]
\centering
\caption{Consistency, solvability, and structural metrics of the released
cases, with the same-footprint synthetic ACTIVSg10k (western U.S.) and
ACTIVSg70k (eastern U.S.) cases as references.}
\label{tab:validation}
\footnotesize
\setlength{\tabcolsep}{1.4pt}
\begin{tabular}{@{}lccccccccc@{}}
\hline
& \textbf{Eastern} & \textbf{Western} & \textbf{MIOHIN} & \textbf{Texas} &
\textbf{Florida} & \textbf{Wisc.} & \textbf{N. Eng.} &
\textbf{10k} & \textbf{70k}\\
\hline
AC PF from stored state          & yes & yes & yes & yes & yes & yes & yes & -- & --\\
Max $|\Delta V|$ vs PYPOWER (pu)  & 0.0075 & 0.0054 & 0.0353 & 0.0075 & 0.0001 & 0.0001 & 0.00001 & -- & --\\
Voltage range (pu)                           & 0.905--1.050 & 0.918--1.056 & 0.933--1.050 & 0.913--1.027 & 0.917--1.020 & 0.904--1.090 & 0.986--1.067 & -- & --\\
Branches above 100\,\% (max \%)              & 1 (100.0) & 0 (99.4) & 10 (140.7) & 0 (99.3) & 0 (98.6) & 2 (126.2) & 0 (89.3) & 0 (86.0) & 0 (98.3)\\
AC losses (\% of load)                       & 1.82 & 3.24 & 1.48 & 1.51 & 1.43 & 3.54 & 0.63 & 1.64 & 3.08\\
Branches per bus                             & 1.607 & 1.607 & 1.672 & 1.570 & 1.604 & 1.480 & 1.356 & 1.271 & 1.260\\
Mean/max node degree                       & 3.21/17 & 3.21/30 & 3.34/17 & 3.14/12 & 3.21/16 & 2.96/11 & 2.71/9 & 2.54/20 & 2.52/23\\
Node-degree fit $R^2$                        & 0.953 & 0.954 & 0.914 & 0.916 & 0.918 & 0.896 & 0.756 & 0.975 & 0.984\\
Fit-area diff.\ vs EI/WECC (\%)        & 13.6/5.8 & 34.3/19.6 & 41.1/28.0 & 57.6/48.1 & 58.9/49.7 & 64.5/56.5 & 85.5/82.3 & 14.7/4.3 & 1.2/20.9\\
Bridges (\% of branches)                     & 7.8 & 9.3 & 4.2 & 9.1 & 9.4 & 14.7 & 12.1 & 31.3 & 28.3\\
Parallels merged: degree/bridges (\%)\textsuperscript{a} & 2.75/10.2 & 2.78/11.5 & 2.89/4.9 & 2.70/12.4 & 2.78/12.0 & 2.65/16.4 & -- & 2.44/32.6 & 2.38/30.0\\
Realism criteria met \cite{Birchfield_2017_MetricBasedValidation}\textsuperscript{b} & 6/13 & 5/13 & 8/13 & 5/13 & 7/13 & 3/8 & -- & 12/13 & 10/13\\
\hline
\end{tabular}
\par\vspace{3pt}
\parbox{\textwidth}{\footnotesize 10k, 70k: ACTIVSg10k and ACTIVSg70k,
from the stored state of their .m files. Branch loading in percent of rating~A,
on PowerWorld's Amp/MVA basis for the released cases (branches the release
files monitor) and on an MVA basis for
ACTIVSg; New England: base case. Losses from the solved state in each .m file
(generation minus load and DC-line losses), checked against the sum of the
branch losses. Branches per bus counts every branch in the file; the other
structural metrics use in-service branches. \textsuperscript{a}In-service branches with parallel circuits merged and
isolated buses removed; the real EI, WECC, and ERCOT networks have mean degrees
of 2.53, 2.40, and 2.45 on this basis
\cite{CotillaSanchez_2012_ComparingTopologicalElectrical}.
\textsuperscript{b}Whole-case criteria 1 to 11 of
\cite[Table~6]{Birchfield_2017_MetricBasedValidation} that the files support
(Wisconsin has no substation data).}
\end{table*}

\subsection{Structural Realism}\label{subsec:realism}
We measure structural realism with the graph metrics of
\cite[Ch.~2]{Snodgrass_2021_Dissertation}, computed on the in-service branches
of each .m file. The contiguous United States has only three interconnections, so a
synthetic case cannot be tested against a sample of real networks; the
comparison measures closeness instead
\cite{CotillaSanchez_2012_ComparingTopologicalElectrical,
Birchfield_2017_MetricBasedValidation}. The degree of a bus is the number of
branches connected to it. Following \cite{Snodgrass_2021_Dissertation}, we fit
a line to the logarithm of the fraction of buses at each degree, from degree~1 to the last degree before the first empty one, and compare the
area under that line with the same area for the real interconnection. The real
Eastern (EI) and Western (WECC) references come from the analysis in \cite{Snodgrass_2021_Dissertation} of their FERC Form~715 planning cases; only their fit areas survive,
not their per-degree counts. Their footprints also differ from the synthetic
ones: the real EI excludes ERCOT, which the Eastern case contains, and the real
WECC extends into Canada and Mexico.

\begin{figure*}[!t]
\centering
\includegraphics[width=\textwidth]{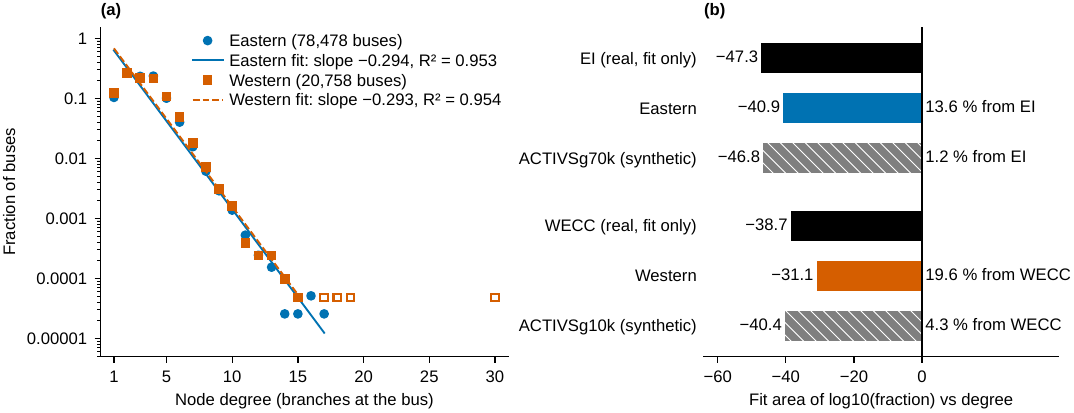}
\caption{Node-degree distributions of the Eastern and Western cases. (a)
Fraction of buses by node degree (isolated buses excluded), with each case's
least-squares line of $\log_{10}$(fraction) against degree, fitted from
degree~1 to the last degree before the first empty one (Eastern 1 to 17, Western 1 to 15; open
markers lie outside the fit). (b) Area under the fitted line for the two
cases, for the real EI and WECC networks (available only as fit areas), and
for the synthetic ACTIVSg70k and ACTIVSg10k cases on the same footprints.}
\label{fig:nodedeg}
\end{figure*}

Fig.~\ref{fig:nodedeg} and Table~\ref{tab:validation} give the results. The
Eastern and Western cases have nearly identical degree distributions (slopes
of $-0.294$ and $-0.293$, with $R^2$ of 0.953 and 0.954). The Eastern fit area
lies 13.6\,\% from the EI value, and the Western fit area 19.6\,\% from the
WECC value. The fit area depends on network size (the synthetic ACTIVSg cases of 200,
500, and 2{,}000 buses lie 81.9, 72.1, and 35.3\,\% from the EI value), so we
use it only as a secondary measure.  Bridges, the
branches whose outage islands part of the network, make up 4.2 to 14.7\,\% of
in-service branches, and in the six larger cases 97 to 99\,\% of them are
69~kV branches, most of them radial taps.

The ACTIVSg10k (western U.S.) and ACTIVSg70k (eastern U.S.) cases of the
graph-statistics family cover the same footprints as the Western and Eastern
cases \cite{Kunkolienkar_2024_TestCaseRepo}, and Table~\ref{tab:validation}
gives their metrics from the same scripts. Counted with every circuit, the
Western and Eastern cases have a mean degree of 3.21 against 2.52 to 2.54. The
like-for-like comparison merges parallel circuits, as the published real-grid
values do \cite{CotillaSanchez_2012_ComparingTopologicalElectrical}. On that
basis, the Eastern and Western cases have mean degrees of 2.75 and 2.78,
8.7\,\% above the real EI (2.53) and 16.0\,\% above the real WECC (2.40), while
ACTIVSg70k lies 5.9\,\% below the EI and ACTIVSg10k 1.8\,\% above the WECC.
Parallel circuits, most of them pairs of transformers in one substation,
account for 53 to 67\,\% of the excess; the rest is topology. On
the whole-case criteria of \cite{Birchfield_2017_MetricBasedValidation}, the
EPIGRIDS cases meet 5 to 8 of 13 (Wisconsin, without substation data, 3 of
8) and the ACTIVSg cases 10 and 12. They fall
outside the real ranges because nearly every unit is in service (97 to
100\,\% against 60 to 80\,\%), installed capacity is up to 1.84 times load
(against 1.2 to 1.6), and substations hold 1.27 to 1.38 buses (against 1.7 to
3.5). Their transformer impedances and their 138~kV line lengths relative to
the minimum spanning tree lie inside the real ranges. The EPIGRIDS cases are
therefore structurally different from the graph-statistics cases, but on these
measures they are denser, and further from the real grids, than those cases.

\subsection{Electrical Realism}\label{subsec:losses}
In their solved base cases, Western and Wisconsin lose 3.24 and 3.54\,\% of
load, Eastern, MIOHIN, Texas, and Florida lose 1.43 to 1.82\,\%, and New
England loses 0.63\,\%. The stored states of ACTIVSg10k and ACTIVSg70k lose
1.64 and 3.08\,\%, so neither family has lower losses on both footprints. We computed each value
two ways, as generation minus load and DC-line losses, and as the sum of the
branch losses from the solved voltages; the two agree within 2~MW in every
case.

Table~\ref{tab:kv} compares the voltage structure of the two interconnection
cases with the real networks. Both cases are heavier at 69~kV and lighter at
138 and 230~kV than the real networks:
56.8\,\% of Eastern lines are at 69~kV
against 42.6\,\% in the EI, and 67.6\,\% of Western lines against 31.9\,\% in
the WECC. The analysis in \cite{Snodgrass_2021_Dissertation} found the same disagreement in the 69 and 138~kV
bus shares of its regional networks and named a remedy in the bus-voltage step
\cite[Ch.~6]{Snodgrass_2021_Dissertation}. The released cases keep it, because
changing the voltage step would make a different network from the one released
in 2022 and carried in PGLib-OPF. The extra 69~kV branches are mostly radial
taps, so they also set the bridge share of Table~\ref{tab:validation} and the
islanding contingencies of Section~\ref{sec:studies}-\ref{subsec:n1}.

\begin{table}[!t]
\centering
\caption{Share of lines and buses by voltage class (kV) in the Eastern and
Western cases and in the real Eastern (EI) and Western (WECC)
Interconnections, in percent.}
\label{tab:kv}
\footnotesize
\setlength{\tabcolsep}{3.5pt}
\begin{tabular}{@{}llrrrrrr@{}}
\hline
\textbf{Network} & & \textbf{69} & \textbf{138} & \textbf{230} &
\textbf{345} & \textbf{500} & \textbf{765}\\
\hline
Eastern     & lines & 56.8 & 36.0 & 3.1 & 3.0 & 1.0 & 0.04\\
EI (real)   & lines & 42.6 & 44.6 & 8.9 & 2.9 & 0.7 & 0.2\\
Western     & lines & 67.6 & 27.6 & 2.5 & 1.3 & 1.0 & --\\
WECC (real) & lines & 31.9 & 47.4 & 17.0 & 1.2 & 2.4 & --\\
\hline
Eastern     & buses & 58.7 & 34.7 & 2.9 & 2.7 & 0.9 & 0.04\\
EI (real)   & buses & 53.2 & 36.7 & 7.2 & 2.1 & 0.5 & 0.1\\
Western     & buses & 68.5 & 25.9 & 3.0 & 1.4 & 1.2 & --\\
WECC (real) & buses & 50.8 & 34.8 & 11.6 & 1.1 & 1.7 & --\\
\hline
\end{tabular}
\end{table}

\subsection{Tools and Reproducibility}\label{subsec:repro}
PowerWorld Simulator (build of June~24, 2026) \cite{PowerWorld_2026_Simulator}
ran the AC power flows and the $N\!-\!1$ analysis, driven through SimAuto with
the esapp Python package, version 0.2.1 \cite{Lowery_2026_ESApp}.
PYPOWER 5.1.21 \cite{Lincoln_PYPOWER} gave the independent power flows, Gurobi 13.0.3
\cite{Gurobi_Manual} the DC optimal power flows, and NetworkX 3.7
\cite{Hagberg_2008_NetworkX} the graph metrics. The OPF benchmark of
Section~\ref{sec:studies}-\ref{subsec:opf} used PowerModels.jl 0.21.7
\cite{Coffrin_2018_PowerModels} with Ipopt 3.14.19 \cite{Wachter_2006_Ipopt}
and the MUMPS linear solver. The Python scripts that ran these checks were
written with Claude \cite{Anthropic_2026_ClaudeOpus48} and checked by the
authors.

\section{Benchmark Studies and Intended Use}\label{sec:studies}
This section reports baseline studies a user can repeat on the released
files, then their limits and intended use.

\subsection{Full AC $N\!-\!1$ Contingency Analysis}\label{subsec:n1}
We ran a full AC $N\!-\!1$ contingency analysis of every case in PowerWorld
(Table~\ref{tab:n1}). The contingency list holds every closed line and
transformer and every in-service unit, 132{,}788 contingencies in Eastern.
Each case uses its own default limit set (PowerWorld's ``Default''): rating~A
in the base case and rating~C after the outage on the Amp/MVA basis (Wisconsin
and New England: rating~A on an MVA basis). Bus voltage limits are those stored in the case: in the .m files, 0.90 to
1.10~pu at every bus of the six larger cases, except a 0.98~pu minimum at 78
Wisconsin generator buses. Loads keep the model of the .pwb, constant power for MW and
constant impedance for Mvar. A violation counts as new when its element is not violated in the solved base
case. The release .pwb files of Eastern, MIOHIN, Texas, and Florida leave every
69~kV line unmonitored; we monitored every branch. Buses
below 0.5~pu are counted as islanded or collapsed rather than as voltage
violations, which is why several worst voltages in Table~\ref{tab:n1} lie just
above 0.5~pu. Between 0 and 22 contingencies per case fail to solve: 18 of these 51 island
part of the network, and 33 do not converge. Every islanding contingency is a bridge
of Section~\ref{sec:validation}-\ref{subsec:realism}, so in the six larger
cases islanding comes almost entirely from 69~kV radial taps, not from the bulk
network. A bridge that does not appear as an islanding contingency either did
not solve, leaves an island with its own generation, or isolates a bus with no
load. New branch overloads of
at least 100\,\% follow 0.5 to 4.2\,\% of the contingencies. In Eastern, one
branch dominates that count: 20422-21190 \#1 sits at 100.0\,\% in the base
case and crosses 100\,\% in 3{,}253 contingencies, mostly unit outages whose
lost output the rest of the system picks up through automatic generation
control. Without it, 2{,}169 Eastern contingencies cause a new overload. In the
four cases with 69~kV lines unmonitored in the release, the worst new overloads
are 69~kV lines, at 210 to 433\,\% after the outage of the transformer that
feeds their pocket. The Eastern run took 248 instance-hours of PowerWorld time (the run times of up to 20 parallel SimAuto instances, summed), a guide to the cost of repeating
it.

\begin{table}[!t]
\centering
\caption{Full AC $N\!-\!1$ contingency analysis of the released cases
(PowerWorld; loading in percent of rating~C on the Amp/MVA basis, and of
rating~A on an MVA basis for Wisconsin and New England).}
\label{tab:n1}
\footnotesize
\setlength{\tabcolsep}{1.5pt}
\begin{tabular}{@{}lrrrrrrr@{}}
\hline
& \textbf{East.} & \textbf{West.} & \textbf{MIOH.} & \textbf{Tex.} &
\textbf{Fla.} & \textbf{Wis.} & \textbf{N.E.}\\
\hline
Contingencies & 132{,}788 & 35{,}517 & 17{,}725 & 12{,}203 & 9{,}546 & 2{,}541 & 381\\
\quad lines & 88{,}369 & 24{,}852 & 12{,}045 & 8{,}058 & 6{,}429 & 1{,}902 & 248\\
\quad transformers & 37{,}646 & 8{,}491 & 4{,}966 & 3{,}461 & 2{,}643 & 560 & 91\\
\quad units & 6{,}773 & 2{,}174 & 714 & 684 & 474 & 79 & 42\\
Unsolved & 22 & 7 & 1 & 5 & 6 & 10 & 0\\
Islanding & 9{,}725 & 3{,}077 & 706 & 1{,}049 & 852 & 360 & 38\\
New overload $\geq$100\,\% & 5{,}208 & 231 & 537 & 219 & 50 & 95 & 16\\
\quad without top branch & 2{,}169 & 219 & 408 & 185 & 45 & 72 & 14\\
New overload $\geq$105\,\% & 1{,}039 & 163 & 256 & 125 & 35 & 62 & 10\\
New voltage violation & 491 & 105 & 25 & 93 & 36 & 142 & 10\\
Worst loading (\%) & 433 & 391 & 212 & 210 & 231 & 248 & 267\\
Worst voltage (pu) & 0.503 & 0.512 & 0.803 & 0.513 & 0.589 & 0.501 & 0.558\\
\hline
\end{tabular}
\par\vspace{3pt}
\parbox{\columnwidth}{\footnotesize Columns: Eastern, Western, MIOHIN,
Texas, Florida, Wisconsin, and the New England base case. Every islanding
contingency is a graph bridge; overloads and voltage violations count new ones
only (not violated in the base case), with every branch monitored.}
\end{table}

\subsection{Optimal Power Flow Benchmark}\label{subsec:opf}
With the tools of Section~\ref{sec:validation}-\ref{subsec:repro}, at
PGLib-OPF's tolerance of $10^{-6}$, we benchmarked optimal power flow on all
seven cases as released in seven formulations: AC polar (ACP), AC rectangular
(ACR), current-voltage (IVR), DC (DCP), linear-programming AC (LPAC),
second-order cone (SOC), and quadratic convex (QC)
\cite{Coffrin_2018_PowerModels} (Table~\ref{tab:opf}). Every case except the
New England base case solves in ACP. That base case commits 11{,}651~MW of
minimum generation against 11{,}324~MW of load, so it needs a unit commitment before an OPF; in its stored dispatch, 21 units
run below their minimum output. Its 12 operating points solve in ACP except Spring13Tight, at
310.6 to 776.9 thousand \$/h.  The SOC and QC relaxations lie
0.08\,\% (Florida) to 0.98\,\% (Eastern) below ACP and agree with each other
within 0.01 percentage points. The release files carry no angle-difference limits (PowerModels then applies
$\pm 60^{\circ}$), which likely explains why QC adds nothing to SOC. ACR and IVR match ACP within 0.01\,\%, except on Western,
where ACR finds a local optimum 0.81\,\% higher, and on Texas, where IVR is
locally infeasible. The PYPOWER AC OPF matches ACP on Texas (1{,}381{,}066 against 1{,}381{,}058~\$/h), and, as a sensitivity, the true unit limits from the .pwb lower the Western
ACP objective by 2.0\,\% and change those of Eastern, MIOHIN, and Texas by
less than 0.01\,\%
(Section~\ref{sec:cases}-\ref{subsec:formats}).

PGLib-OPF v23.07 already holds five of these cases (Florida, Texas, MIOHIN,
Western, and Eastern) \cite{PGLibOPF_2023_v2307}. Its versions
limit every branch angle difference to $\pm 30^{\circ}$ and replace the cost
curves of Florida, Texas, and Eastern, which raises the Texas ACP objective by
36\,\%. Loads, voltage and generator limits, and branch impedances and
ratings are unchanged.
Our runs on the PGLib
files reproduce its published baseline to four or five significant digits,
which checks our setup. Users comparing methods against published PGLib-OPF
results should use the PGLib files; users who need the cases as distributed,
with the power flow, $N\!-\!1$, and load series of this paper, should use the
release files.

\begin{table}[!t]
\centering
\caption{Optimal power flow benchmark of the released cases with
PowerModels.jl and Ipopt (objectives in thousands of \$/h; relaxation gaps
below ACP in \%), with the ACP objective of the PGLib-OPF v23.07 file of the
same case.}
\label{tab:opf}
\footnotesize
\setlength{\tabcolsep}{2.5pt}
\begin{tabular}{@{}lrrrrrrr@{}}
\hline
\textbf{Case} & \textbf{Buses} & \textbf{ACP} & \textbf{DCP} & \textbf{LPAC} &
\begin{tabular}[t]{@{}c@{}}\textbf{SOC}\\ \textbf{gap}\end{tabular} & \begin{tabular}[t]{@{}c@{}}\textbf{QC}\\ \textbf{gap}\end{tabular} &
\begin{tabular}[t]{@{}c@{}}\textbf{PGLib}\\ \textbf{ACP}\end{tabular}\\
\hline
Eastern & 78{,}478 & 12{,}415.2 & 12{,}284.2 & 12{,}403.5 & 0.98 & 0.98 & 15{,}315.9\\
Western & 20{,}758 & 2{,}617.9 & 2{,}568.2 & 2{,}611.9 & 0.36 & 0.35 & 2{,}618.6\\
MIOHIN & 10{,}189 & 1{,}686.9 & 1{,}665.6 & 1{,}679.6 & 0.85 & 0.85 & 1{,}686.9\\
Texas & 7{,}336 & 1{,}381.1 & 1{,}361.1 & 1{,}378.6 & 0.20 & 0.20 & 1{,}882.4\\
Florida & 5{,}658 & 1{,}226.8 & 1{,}219.8 & 1{,}225.2 & 0.08 & 0.08 & 1{,}207.3\\
Wisconsin & 1{,}664 & 212.9 & 208.2 & 212.0 & 0.41 & 0.41 & --\\
\begin{tabular}[t]{@{}l@{}}New\\ England\\ (base)\end{tabular} & 250 & infeas.\textsuperscript{a} & infeas.\textsuperscript{a} & 424.3\textsuperscript{b} & -- & -- & --\\
\hline
\end{tabular}
\par\vspace{3pt}
\parbox{\columnwidth}{\footnotesize Buses: in-service buses.
\textsuperscript{a}Locally infeasible: minimum generation exceeds load.
\textsuperscript{b}The LPAC cosine relaxation absorbs the excess generation as
losses. PGLib: our run on the v23.07 file, which matches PGLib's published
baseline.}
\end{table}

\subsection{Limitations and Intended Use}\label{subsec:limits}
A user should know eight properties of the cases as released. First, the New
England base case commits more minimum generation than load, so its 12
operating points are the ones to use for OPF. Second, MIOHIN, Wisconsin, and the 69~kV lines of Eastern keep base-case overloads
(Section~\ref{sec:validation}-\ref{subsec:solve}), and a MATPOWER-style DC OPF
of MIOHIN needs 21.4~MW of branch-limit relief. Third, the cases
are not hardened against branch outages
(Section~\ref{sec:studies}-\ref{subsec:n1}); \cite[Ch.~9]{Snodgrass_2021_Dissertation} lists that hardening as future work. Fourth, the MATPOWER files fix the
limits of units off automatic generation or voltage control at their dispatch,
while the .RAW and .pwb hold the true limits
(Section~\ref{sec:cases}-\ref{subsec:formats}). Fifth, 286 Eastern units, 89
of them nuclear, carry zero cost as released. Sixth, Texas, Florida, and MIOHIN
are sub-networks of the Eastern case
(Section~\ref{sec:construction}-\ref{subsec:extract}). Seventh, the cases carry more 69~kV and fewer 230~kV lines than the real
networks, are denser than them even with parallel circuits merged, and commit
nearly every unit (Section~\ref{sec:validation}-\ref{subsec:realism}).
Eighth, the release holds no dynamic models, the line parameters are not
compared here with measured values, and
the load series is solved only at each case's stored hour.

Within those limits, the cases suit AC power flow, contingency, and OPF
benchmarking from state to interconnection scale, and, being structurally
different from the graph-statistics cases, they add a second built-from-scratch
family to the training and test data of machine-learning models.

\section{Conclusion}\label{sec:conc}
This paper documented and validated the seven EPIGRIDS cases, from 250 to
78{,}484 buses, as released under the CC~BY~4.0 license
\cite{WISPOPOP_2022_SyntheticModels, Snodgrass_2026_EPIGRIDSZenodo}. They were
built with the KCL-only construction code of
\cite{Snodgrass_2021_Dissertation, Snodgrass_2026_GridForge} instead of the
graph-statistics approach behind most large public synthetic grids, so they add
to the variety of construction methods available for benchmarking power system
methods and for training machine-learning models.

The checks support that use: the formats agree at the solved operating
point, every case solves an AC power flow, a full AC $N\!-\!1$ completes on
every case, and all but the New England base case solve an AC optimal power
flow. On the same footprints, the Eastern and Western cases are structurally
different from the ACTIVSg cases but denser than the real networks, while the
ACTIVSg cases lie closer to the real networks and meet more published realism
criteria.

This paper enables three lines of future work. The first trains AI models on a
single family of synthetic grids and on several, to measure how a model trained
on one family performs on another and how many families give the best results
before the returns diminish or the variety starts to hurt. It would also measure how many members
of each family, and how many variants and scenarios of each member, the
training needs. The second compares the public synthetic grid families in
detail, with two-sample statistics of their degree, path-length, and
line-parameter distributions and the hourly solvability of their load series.
The third is a successor construction method that enforces Kirchhoff's voltage
law within the optimization \cite{Snodgrass_2021_Dissertation}.

\section*{Data Availability}
The seven EPIGRIDS cases are openly available under the Creative Commons
Attribution 4.0 International (CC~BY~4.0) license in the \emph{Synthetic
Electric Network Models} repository \cite{WISPOPOP_2022_SyntheticModels},
archived on Zenodo \cite{Snodgrass_2026_EPIGRIDSZenodo}, and listed on the
Texas A\&M Electric Grid Test Case Repository \cite{TAMU_2026_EPIGRIDSTestCases}.
The companion load series of Section~\ref{sec:cases}-\ref{subsec:load} is
archived on Zenodo under the same license
\cite{Snodgrass_2026_LoadSmithZenodo}. The regional networks and evaluation
scripts of the construction method are archived separately
\cite{Snodgrass_2026_GridForgeZenodo}. Users of the cases should cite this
paper, the archived release \cite{Snodgrass_2026_EPIGRIDSZenodo}, and the paper
that describes the construction method \cite{Snodgrass_2026_GridForge}.

\section*{Acknowledgment}

The authors thank Andrew Bryce, Anna Iwanski, Caleb Kuske, Tyler Whitmore, and
Jacob Yatso, then undergraduate students at the University of
Wisconsin-Madison, for their work on the Eastern and Western cases in
PowerWorld.

The authors used Claude (Anthropic) \cite{Anthropic_2026_ClaudeOpus48} for
copy-editing this manuscript, and to write the Python scripts that summarized
and analyzed the released cases for Sections~\ref{sec:construction}
to~\ref{sec:studies}. The authors checked those scripts and their results. AI
was not used to create the cases or the code that created them; the cases and
their construction code are solely the work of the authors, and the words of
this paper are the first author's own.

The authors declare no competing financial or non-financial interests.

\bibliographystyle{IEEEtran}
\bibliography{references-ArXiv}

\end{document}